\documentstyle[aps,preprint]{revtex}
\begin{document}
\draft
\title{Static Friction Phenomena in Granular Materials: Coulomb Law vs.
Particle Geometry}
\author{Thorsten P\"oschel and Volkhard Buchholtz}
\address{
	HLRZ, KFA J\"ulich, Postfach 1913,
	D--52425 J\"ulich, Germany;\\
	Humboldt--Universit\"at zu Berlin,
	FB Physik, Institut f\"ur Theoretische Physik,
	Unter den Linden 6, D--10099 East--Berlin
}
\date{\today}
\maketitle
\begin{abstract}
The static as well as the dynamic behaviour of granular material are determined
by dynamic {\it and} static friction. There are well known methods to include
static friction in molecular dynamics simulations using scarcely understood
forces. We propose an Ansatz based on
the geometrical shape of nonspherical particles which does not involve an
explicit expression for static friction. It is shown that the simulations
based on this model are close to experimental results.
\end{abstract}
\setcounter{topnumber}{3}
\renewcommand{\topfraction}{1.0}
\setcounter{topnumber}{5}
\setcounter{bottomnumber}{5}
\setcounter{totalnumber}{5}
\renewcommand{\topfraction}{1}
\renewcommand{\bottomfraction}{1}
\renewcommand{\textfraction}{0}
\narrowtext
The behaviour of fluidized dry granular material, like sand or powder, reveals
a rich variety of effects, which can not be observed in other substances. Those
effects have been observed and investigated by many scientists over a long
period~\cite{coulomb}--\cite{jaeger_nagel_science}.
Examples for the most interesting effects are
fluidization, convection cells and heap formation under
vibration  \cite{faraday}\cite{evesque}--\cite{gallas_h_s},
size segregation (``Brazil--nut'' effect)
\cite{haff_werner}--\cite{brazil} deformation under shear force~\cite{shear},
shape segregation of differently shaped grains in a pipe~\cite{sokol_herrmann}
and clustering instabilities~\cite{goldhirsch_zanetti}.
Density waves emitted from outlets~\cite{outlets}
inside material flowing through pipes~\cite{pipe} and at the surface of an
inclined chute~\cite{chute} have been intensively investigated.
Recent results gave evidence that convection cells due to walls or amplitude
modulations play an important role in the process of the formation of
macroscopic structures~\cite{gallas_h_s}. Of
particular interest are the dynamic as well as the static behaviour of
avalanches going down the slope of a sand pile. Theoretical as well as
experimental investigations \cite{bak}--\cite{SOC} led to the
hypothesis that their mass and their time distributions can be described by the
self organized criticality--model. There are experiments, however, that do not
agree with this hypothesis~\cite{evesque} \cite{keinSOC}.
Recently many experimental observations have been reproduced by numerical
simulations. There is a wide variety of simulation methods including
Monte--Carlo simulations (e.g.~\cite{devillard}), molecular dynamics
simulations
(e.g.~\cite{cundall_strack}\cite{haff_werner}\cite{sokol_herrmann}),
and random walk approaches~\cite{caram}. These simulations gave
many interesting information on the microscopic effects underlying the
behaviour of macroscopic amounts of granular material.
Many of the effects observed in experiments are consequences of static
friction between the grains. In most of the current simulations special terms
for static friction are used to mimic static behaviour of granular material
e.g.~\cite{cundall_strack}\cite{lee}.
The aim of this paper is to show that it is
possible to reproduce the experimental results by molecular dynamics
simulations without introducing such a static friction force but by simulating
nonspherical particles. We show that our simulations with nonspherical
particles agree better with experimental results than equivalent
simulations introducing static friction forces as it is usually done.
\par
Since it is extremely complicate to calculate collisions of cubic
particles we choose in two dimensions particles
similar to squares but consisting of spheres. A further advantage of this
model is that we are able to vary the shape steadily from a sphere almost to a
square. A related Ansatz for nonspherical grains was recently done by Gallas
and Soko{\l}owski~\cite{gallas_s}, there each grain consists of two spheres
rigidly glued to each other.
Each of our nonspherical particles $k$ consists of four spheres with
equal radii $r_{i}^{(k)}$, located at the edges of a square of size $L^{(k)}$,
and one sphere with radius
$r_{m}^{(k)} = L^{(k)} / \sqrt{2}- r_{i}^{(k)} $ in the
middle of the square (fig.~\ref{geopart}). For the case that two spheres $i$
and $j$ of the same particle $k$ or of different particles touch each other
during a collision there acts the force
\begin{eqnarray*}
\vec{F}_{ij}^{C} = \Big[ Y (r_i+r_j-|\vec{x}_i-\vec{x}_j|)+\gamma
	 m_{eff} |\dot{\vec{x}}_i - \dot{\vec{x}}_j| \Big]
	\frac{\vec{x}_i - \vec{x}_j}{|\vec{x_i}-\vec{x}_j|}
\end{eqnarray*}
\begin{eqnarray*}
\mbox{with \hspace{0.3in}}
m_{eff} = \frac{m_i \cdot m_j}{m_i + m_j}
\end{eqnarray*}
where $Y$ is the Young modulus and $\gamma$ is the phenomenological friction
coefficient. In addition to
the forces between each two particles of the system, there are forces
between each pair of spheres $i,j$ where $i$ and $j$ both belong to the
same grain, due to a
damped spring
\begin{eqnarray*}
\vec{F}_{ij}^{S} = \Big[ \alpha (C^{(k)} - |\vec{x}_i-\vec{x}_j|) +
	\gamma_{Sp}\frac{m_i}{2} |\dot{\vec{x}}_i -
\dot{\vec{x}}_j |
	\Big]
	\frac{\vec{x}_i - \vec{x}_j}{|\vec{x_i}-\vec{x}_j|}
\end{eqnarray*}
where $\alpha$ and $\gamma_{Sp}$ are the spring constant and the damping
coefficient. If the spheres $i$ and $j$ are both located at the same edge of
the
square then $C^{(k)}$ equals $L^{(k)}$, if $i$ is the central sphere then
$C^{(k)} = L^{(k)}/\sqrt{2}$.
\par
The dynamics of large numbers of such nonspherical particles was investigated
simulating the flow of granular material in a rotating cylinder under gravity.
For the integration
we used a sixth order predictor-corrector method
\cite{predictor}. In a cylinder of diameter $D=260$ we simulated the flow of
1000 nonspherical
particles of different size $L^{(k)}$ with Gaussian probability distribution
$p(L^{(k)})=N(3,1)$
each consisting of five spheres.  For the
parameters we chose $Y=10^4~kg/s^2$, $\gamma=1.5 \cdot 10^4~s^{-1}$,
$\alpha=10^4~kg/s^2$,
$\gamma_{Sp}=3 \cdot 10^4~s^{-1}$ and $r_i^{(k)}=1/4 \cdot L^{(k)}$. The
cylinder consists of
spheres with different radii to mimic a rough surface. The mean value of these
spheres equals the mean value of the $L^{(k)}$. The cylinder was rotated
clockwise with very low angular velocity $\Omega$.
During the uniform rotation of the cylinder the flow of the grains was very
inhomogeneous, due to avalanches going down the
inclined surface. This behaviour is called stick--slip motion.
The time evolution of the slope $\Theta$
of the surface as well as the averaged velocity $\bar{v}$ of the particles at
the surface for a fixed angular velocity $\Omega=0.002~s^{-1}$ are
drawn in fig.~\ref{winkel_v_t} (curves $v(a), \Theta (a)$). The angle was
plotted in rad, $\bar{v}$ in $50\cdot s^{-1}$. Since the
number of particles is not too large our surfaces are not smooth. Hence we have
to determine the inclination indirectly as the angle between the straight line
connecting the centre of mass point of the grains and the middle point of the
rotating cylinder and the direction of gravity.
The angle and particularly the average velocity of the
surface particles fluctuate drastically and irregularly as it is
typical for stick--slip motion.
This behaviour was observed experimentally before by Briscoe, Pope and
Adams~\cite{briscoe_pope_adams} and by Rajchenbach~\cite{rajchenbach_prl}.
The plots $v(b), \Theta (b)$ in the same figure show the equivalent data for
the simulation using spherical particles. The radii of the spheres were
Gauss-distributed too with $p(r_i)=N[1,1]$.
The spherical grains undergo the same
force as the spheres of which the nonspherical particles consist.
To mimic static friction we include for the case of spherical particles
rotation as a further degree of freedom of the grains and add the force
\begin{eqnarray*}
\vec{F}_{ij}^{sf}=\min \{- \gamma_s m_{eff} |\vec{v}_{rel}| , \mu
|\vec{F}_{ij}^{C}| \} \left({0 \atop 1} ~{-1 \atop 0} \right)
                  \frac{\vec{x}_i - \vec{x}_j}{|\vec{x_i}-\vec{x}_j|}
\end{eqnarray*}
\begin{eqnarray*}
\mbox{with \hspace{0.3in}}
\vec{v}_{rel} = (\dot{\vec{x}}_i - \dot{\vec{x}}_j) + r_i \cdot
\dot{\vec{\omega}}_i +
	r_j \cdot \dot{\vec{\omega}}_j \hspace{0.5cm} ,
\end{eqnarray*}
where $\dot{\vec{\omega}}_i$ is the angular velocity of the $i$-th particle,
$\gamma_s$ is the shear friction coefficient, and $\mu$ is the Coulomb
parameter ($\gamma_s = 3 \cdot 10^4~s^{-1}$, $\mu = 0.5$). This Ansatz is the
most popular to include static friction between particles which roll on each
other into the expressions for the forces used in molecular dynamics
simulations. It was introduced by Cundall and Strack~\cite{cundall_strack} and
modified by Haff and Werner~\cite{haff_werner}. Most of the molecular dynamics
simulations of granular material base on this Ansatz. The force
$\vec{F}_{ij}^{sf}$ was implemented only for the simulation of spherical grains
but not for the nonspherical.
\par
Obviously the qualitative behaviour of the slope $\Theta$ resemble each other
in both simulations but quantitatively we get for nonspherical grains more than
twice the mean angle ($\overline{\Theta_{ns}}$) than for spherical
($\overline{\Theta_{sp}}$). For very low rotation velocity $\Omega = 2 \cdot
10^{-3}$ we found $\overline{\Theta_{sp}}=7^o$ and
$\overline{\Theta_{ns}}=19^o$.
In the experiment~\cite{rajchenbach_prl} was measured $\Theta \approx 27^o$.
The average velocity of the surface grains differs significantly too for
both cases. The typical avalanches in the case of nonspherical particles can
not be observed for spheres. The curve $\bar{v}$ (b) is much smoother.
In the experiment one observes stick--slip motion~\cite{rajchenbach_prl}.
Fig.~\ref{theta_omega} shows the slope $\Theta$ of the surface as a function of
the angular velocity of the cylinder $\Omega$ for both nonspherical
and spherical grains. In both cases the curves are close to
a straight line. For much higher angular velocities than used in our
simulations the grains do not move
stick--slip like but continuously. In this regime was experimentally found
$\Omega \sim
(\Theta - \Theta_c)^m$, with $m=0.5 \pm 0.1$~\cite{rajchenbach_prl}. With the
same Ansatz we find $m \approx 1.25$ for the stick--slip regime.
As shown above the simulation with nonspherical grains coincides much better
with the
experimental observations than equivalent simulations using spheres.
\par
In our second simulation we investigate the evolution of a stable pile of
granular material by continuously dropping particles on the top of the pile due
to the experiment of Held et.~al.~\cite{held}.
Beginning with an empty rough plane we drop the next particle when the maximum
velocity vanishes, i.e.~when it is smaller than a given very small threshold
$v_{max} \ll 1$. The rough
plane was simulated by a chain of fixed spheres of random radii with mean
$\overline{r_i} = \overline{L^{(i)}}$ where $L^{(i)}$ is the size of the
$i$--th nonspherical grain. The parameters
of the simulation were the same as in the previous experiment.
\par
During the simulation we noticed that the slope of a pile of nonspherical
grains does not
depend on the number of particles.
For spherical grains, however, the heap dissolves under gravity, i.e. the slope
decreases with increasing particle number.
There are molecular dynamics simulations of stable piles with spherical
grains,
e.g.~\cite{lee}, but there the particles are not allowed to roll on each other,
hence they can only slide, this behaviour does not correspond
to experimental reality.
\par
If the platform above which the heap is built up has a finite length $P$ one
can investigate the fluctuations
of the mass $m_h$ of a heap of definite size and the distribution of the
size of the avalanches, i.e.~the mass fluctuations of the heap.
In fig.~\ref{masse_zeit} is drawn the time series of the
mass $m_h$ for fixed $P$. The mass fluctuates irregularly due to
avalanches of different size going down the surface of the heap.
The size distribution of the avalanches follow a power law,
fig.~\ref{log_log_size} shows the log--log--plot of the spectrum. For the
exponent $h(N_A) \sim (N_A)^m$ we found $m\approx -1.4$. The experiments yield
$m\approx -2.5$ \cite{held} and $m\approx -2.134$ \cite{bretz}.
For the case of spherical grains we cannot find avalanches.
\par
The ratio between the size of a grain and the radii of the
spheres at the corners determines whether the grains
shape is closer to a sphere or to a square. Hence we define a shape value
\begin{eqnarray*}
S=1-R_{min}^{cc}/R_{max}^{cc},
\end{eqnarray*}
where $R_{min}^{cc}$ and $R_{max}^{cc}$ are the extremal values of the distance
between the convex cover of the nonspherical grain and its central point
(fig.~\ref{winkel_excentr}).
For the limit $S
\rightarrow 0$ the grains have the shape of spheres. The function reaches its
maximum $S_m = 0.255$ for a grain which convex cover is most
similar to that of a square.
To investigate the
influence of the the shape $S$ of the grains on the result of the simulations
we have to scale the density $\rho$ of the material the grains consist of to
ensure that the total mass of each grain remains constant.
Fig.~\ref{winkel_excentr} shows the angle of the heap as a function of the
shape
$S$. For grains with shape $S=S_m$, which corresponds to
$(L^{(k)}/r_i^{(k)})_{S_m} = 9.66$~, the inclination of the heap reaches a
maximum too. The angle $\Phi \approx23.1^o$ agrees with experimental data,
Bretz~et.~al. \cite{bretz} found $\Phi\approx 25^o$. Each other value $S$
corresponds to two different particle shapes both closer to a sphere than
the $S_m$--particle. The values marked by $\odot$ are due to grains with
$L^{(k)}/r_i^{(k)} \le (L^{(k)}/r_i^{(k)})_{S_m}$,
$+$ designates the slope of the heap for particles with
$L^{(k)}/r_i^{(k)} \ge (L^{(k)}/r_i^{(k)})_{S_m}$. As expected the slope $\Phi$
of the heap rises with growing
$L^{(k)}/r_i^{(k)}$ up to $S$ reaches its maximum
$S=S_m$. For larger ratios $L^{(k)}/r_i^{(k)}$ ($S < S_m$) the slope
$\Phi$ decreases. The dashed line in fig.~\ref{winkel_excentr} displays the
inclination $\Phi_{sp}$ we measured for a heap of spheres, which corresponds to
$S \rightarrow 0$. The value $\Phi_{sp}$ gives a lower boundary for the slope.
The observed $\Phi$ values for $S \in (0,S_m)$ lie between $\Phi_{sp}$ and
$\Phi(S_m)$.
The calculation for the data shown in fig.~\ref{winkel_excentr} are
very computer time consuming. For this reason we are not able to present more
data.
\par
The simulations described above demonstrate that nonspherical grains are
able to describe the static behaviour of granular materials, such as
stick--slip motion in a rotated cylinder at very low angular velocity and the
angle of repose of a sandpile. It is shown that equivalent simulations with
spherical grains and
an additional term which describes the static friction
due to the Coulomb law
could neither reproduce the experimental results for stick--slip motion nor for
the angle of repose of a sand pile. The angle of repose reaches its extremal
value for grains which shape is close to a square.
\par
Hence we conclude that our microscopic model of nonspherical grains supplies a
possible description of the static behaviour of a granular material. The
results
regarding non--sphericity agree well with those in \cite{gallas_s}
\par
The authors thank J.~A.~C.~Gallas and H.~J.~Herrmann for stimulating and
enlightening
discussions.

\begin{figure}[p]
\caption{Shape of a nonspherical particle.}
\label{geopart}
\end{figure}
\begin{figure}[p]
\caption{Time evolution of the slope $\Theta$ of the surface and the
averaged velocity $\bar{v}$ of the particles at the surface of the flow for
nonspherical (a) and spherical (b) grains. Due to avalanches $\overline{v_a}$
fluctuates significantly, while in case (b) occur only very small avalanches.}
\label{winkel_v_t}
\end{figure}
\begin{figure}[p]
\caption{The inclination $\Theta$ of the surface as a function of
the angular velocity $\Omega$ of the cylinder.}
\label{theta_omega}
\end{figure}
\begin{figure}[p]
\caption{Total mass of a pile of nonspherical grains on a platform of finite
length $P = 820$.}
\label{masse_zeit}
\end{figure}
\begin{figure}[p]
\caption{Size distribution of the avalanches. The line
displays the function $h(N_A)=(N_A)^{-1.4}$.}
\label{log_log_size}
\end{figure}
\begin{figure}[p]
\caption{Slope $\Phi$ of a heap over the shape value $S$ for grains with
$L^{(k)}/r_i^{(k)} \le (L^{(k)}/r_i^{(k)})_{S_m}$
($\odot$) and $L^{(k)}/r_i^{(k)} \ge
(L^{(k)}/r_i^{(k)})_{S_m}$ ($+$). The dotted line
leads the eye to the function $\Phi = 130 \cdot S + const$. The dashed line
displays the inclination observed in simulation with spherical particles.}
\label{winkel_excentr}
\end{figure}
\end{document}